\documentstyle[12pt]{article}

\begin{document}
\sloppy
\begin{flushright}{KEK-TH-546\\ Oct. '97}\end{flushright}
\vskip 1.5 truecm
\centerline{\large{\bf Domain wall solution for }}
\centerline{\large{\bf vectorlike model}}
\vskip .75 truecm
\centerline{\bf Tomohiro Matsuda
\footnote{matsuda@theory.kek.jp}}
\vskip .4 truecm
\centerline {\it National Laboratory For High Energy Physics (KEK)}
\centerline {\it Tsukuba, Ibaraki 305, Japan}
\vskip 1. truecm
\makeatletter
\@addtoreset{equation}{section}
\def\theequation{\thesection.\arabic{equation}}
\makeatother
\vskip 1. truecm
\begin{abstract}
\hspace*{\parindent}
Domain wall solution for $N_{c}=N_{f}$ supersymmetric QCD
is constructed.
Astrophysical implications of the domain wall configuration
is also discussed.
\end{abstract}
\newpage
\section{Introduction}
\hspace*{\parindent}
When one  extends the validity of the low energy effective field theory
to energy scales much higher than its characteristic mass scale,
one faces
to a scale hierarchy problem.
A typical example is the gauge hierarchy problem of the
Standard Model of the strong and electroweak interactions, seen as
a low-energy effective theory.
When the Standard Model is extrapolated to cut-off scales
$\Lambda\gg 1$Tev, there is no symmetry protecting the mass of
the elementary Higgs field from acquiring large value,
and therefore the masses of the
weak gauge bosons, receive large quantum corrections proportional
to $\Lambda$.
The most popular solution to the gauge hierarchy problem of the
Standard Model is to extend it to a model with global
N=1  supersymmetry, effectively broken at a scale
$M_{Soft}\sim 1$Tev.
These extensions of the Standard Model, for instance the
Minimal Supersymmetric Standard Model(MSSM), can be
safely extrapolated up to cut-off scales much higher than the
electroweak scale, such as the supersymmetric unification
scale $M_{U}\sim 10^{16}$Gev, the string scale
$M_{s}\sim 10^{17}$Gev,
or Planck scale $M_{P}=2.4\times 10^{18}$Gev.
To go beyond  MSSM, one must move to a more fundamental theory
with spontaneous supersymmetry breaking which should be induced
by the dynamical effects to explain the hierarchy.
One of the candidate of the  mechanism of supersymmetry breaking
is the dynamical supersymmetry breaking (DSB) models\cite{Dine}
in which supersymmetry is broken at a low energy scale($\sim 10^{7}GeV$).
A few years ago, a new mechanism for DSB was proposed
by Yanagida et al.\cite{vector} and then applied to many other
models of dynamical symmetry breaking\cite{dvali}.
Unlike other DSB models,  supersymmetry breaking is realized in this model
with a vectorlike gauge theory  and  only one gauge group is required
in the breaking sector.
It is also surprizing that the scale of  gaugino condensation
is not fixed but dynamical in a sense that it explicitly
depends on the vacuum expectation value of the additional singlets.
To be more specific, let us consider the vectorlike supersymmetry
breaking model of ref.\cite{vector}.
Here we consider a supersymmetric SU(2) gauge theory with
four doublet chiral superfields $Q_{i}$ and six singlets
$Z^{ij}=-Z^{ji}$.
Here, $i$ and $j$ denote the flavor indices ($i,j=1,...,4$).
We introduce a tree-level superpotential:
\begin{equation}
  W_{0}=\lambda^{kl}_{ij}Z^{ij}Q_{k}Q_{l}.
\end{equation}
In this model, supersymmetry is not broken classically but
broken by the quantum deformation of the flat direction.
According to Seiberg\cite{exact}, holomorphy implies
a constraint on the gauge-invariant degrees of freedom
$V_{ij}=-V_{ji}\sim Q_{i}Q_{j}$:
\begin{equation}
  PfV_{ij}=\Lambda^{4}.
\end{equation}
With an additional chiral superfield $S$, we can impose this
holomorphic constraint by means of the non-perturbative
superpotential of the form:
\begin{equation}
  W_{dyn}=S(PfV_{ij}-\Lambda^{4})+\lambda_{ij}^{kl}Z^{ij}V_{kl},
\end{equation}
where $PfV_{ij}$ denotes the Pfaffian of the antisymmetric
matrix $V_{ij}$, and $\Lambda$ is a dynamical scale
of the SU(2) gauge interaction.
This effective superpotential yields conditions for supersymmetric
vacua
\begin{eqnarray}
  PfV_{ij}&=&\Lambda^{4}\nonumber\\
  \lambda^{kl}_{ij}V_{kl}&=&0.
\end{eqnarray}
These two conditions do not stand simultaneously.
This mechanism of supersymmetry breaking is very similar to
O'Raifeartaigh model.
The relation to the ordinary supersymmetric QCD theory is
established when  we consider a limit
\begin{eqnarray}
  \lambda^{kl}_{ij}&\rightarrow&0\nonumber\\
  \lambda^{kl}_{ij}Z^{ij}&=&m^{kl}\sim const.
\end{eqnarray}
In this paper we construct a domain wall solution for the
vectorlike DSB sector described above
and discuss its astrophysical implications.
\section{Domain wall configuration for the vectorlike sector}
\hspace*{\parindent}
Let us first consider an $SU(N_{c})$ gauge theory with matter consisting of
$N_{f}$ massive chiral superfields $Q_{i}$ and $\overline{Q_{i}}$
(massive SQCD) for $N_{c}=N_{f}$ and
construct a domain wall solution\cite{shifman}.
Here we consider a classical superpotential of the form:
\begin{equation}
  W_{0}=m^{i}_{j}Q_{i}\overline{Q}^{j}
\end{equation}
The exact effective superpotential of the model may be written
in terms of the gauge-invariant low-energy degrees of freedom
$M^{i}_{j}, B$ and $\overline{B}$.
Because we do not consider terms of the form $bB+\overline{b}\overline{B}$,
expectation values of lower components of chiral superfields is given by
\cite{exact}:
\begin{eqnarray}
  M^{i}_{j}&=&<Q^{i}\overline{Q}_{j}>= \Lambda^{(3N_{c}-N_{f})/N_{c}}
  (det[m^{i}_{j}])^{1/N_{c}}(m^{i}_{j})^{-1}\nonumber\\
  B&=&\overline{B}=0
\end{eqnarray}
The phases from the fractional power $1/N_{c}$ corresponds to $N_{c}$
different ground states which is consistent to the arguments of the
Witten  index.
The exact superpotential is now written by
\begin{equation}
  W_{eff}=S(detM^{i}_{j}-B\overline{B}-\Lambda^{4})+m^{i}_{j}Q_{i}
  \overline{Q}^{j}
\end{equation}
where $S$ is an additional chiral superfield which is
introduced to make the holomorphic constraint manifest.
For $N_{c}=2$, the classical superpotential is written by
\begin{equation}
  W_{0}=m^{ij}Q_{i}Q_{j}
\end{equation}
and the explicit form of the constraint is given by\cite{exact}
\begin{equation}
  V^{ij}=<Q^{i}Q^{j}>=\Lambda^{2}(Pf[m^{ij}])^{1/2}(m^{ij})^{-1}.
\end{equation}
As a result, the exact superpotential of the model is written by
\begin{equation}
  W_{eff}=S(PfV_{ij}-\Lambda^{4})+m^{ij}V_{ij}
\end{equation}
The effective scale $\Lambda_{L}$ of the low-energy SU(2)
along this trajectory is given to all orders by the one-loop
matching of the gauge coupling at the quark mass scale
$m\sim (Pf [m^{ij}])^{1/2}$ and leads \cite{match}
\begin{equation}
  \Lambda_{L}^{6}=m^{2}\Lambda^{4}.
\end{equation}
This relation is consistent to the Konishi anomaly if
$\Lambda_{L}$ is regarded as the scale of gaugino condensation.
In the pure SU(2) gauge theory an
effective superpotential $\sim\Lambda_{L}^{3}$
is generated:
\begin{equation}
  \label{sup}
  W_{eff}=m\Lambda^{2}.
\end{equation}
This effective superpotential may also be obtained by replacing
the matter fields by their vacuum expectation values.
In this model, contrary to the one we have discussed in the previous
section, supersymmetry is not broken but flat direction
is modified by the dynamical effect.
Recently, the domain wall solution for supersymmetric QCD theories are
constructed by Shifman et al\cite{shifman} and found that they are
BPS-saturated.
The key point is that the cental extension is automatically
zero for all spatially localized field configurations but need
not necessary vanish for those field configurations that
interpolate between distinct vacua at spatial infinities (domain wall).
The central charge appears at both classical and at the
one-loop level as a quantum anomaly.
The explicit form is
\begin{equation}
  \left.\{\overline{Q}_{\dot{\alpha}}\overline{Q}_{\dot{\beta}}\}=
  4(\vec{\sigma})_{\dot{\alpha}\dot{\beta}}\int d^{3}x
  \vec{\bigtriangledown}\left\{\left[W-\frac{N_{c}-N_{f}}{16\pi^{2}}
      TrW^{\alpha}W_{\alpha}\right]+t.s.d.\right\}
\right|_{\overline{\theta}=\theta=0}.
\end{equation}
where $t.s.d.$ denotes total superderivatives.
In our model, we should set $N_{c}=N_{f}$ thus we can neglect
the second term.
The first term comes from  the superpotential(\ref{sup})
and it parametrizes  gaugino condensation.
The domain wall configuration for the present model
is almost the same as that was discussed in ref.\cite{shifman}.
The domain wall configuration
appears because $Z_{2N_{c}}$ symmetry is spontaneously
broken by gaugino condensation, and  have the energy
\begin{equation}
  E=2A(W_{*1}-W_{*2})
\end{equation}
where $W_{*i}$ denotes the vacuum expectation value of
the superpotential in the i-th domain.
Now let us further discuss the domain wall configuration
of the vectorlike sector, namely $N_{c}=N_{f}$ supersymmetric
QCD with singlets.
The crucial difference from the ordinary supersymmetric QCD
theory is the presence of
the additional scalar potential of the form:
\begin{equation}
  V_{add}\sim|\lambda V|^{2}
\end{equation}
which breaks  supersymmetry when dynamical effects are included.
To establish the contact with the original supersymmetric theory,
here we consider a limit
\begin{eqnarray}
  \lambda^{kl}_{ij}&\rightarrow&0
\end{eqnarray}
while the lowest component of $\lambda^{kl}_{ij}Z^{ij}$ remains
finite:
\begin{equation}
  Z'^{kl}\equiv \lambda^{kl}_{ij}Z^{ij}.
\end{equation}
In this limit, the vectorlike sector can be regarded as an ordinary
supersymmetric QCD with a mass parametrized by $Z'$.
The domain wall configuration is constructed in the same way.
The energy of the field configuration $E$ now depends on $Z'$:
\begin{equation}
  E\sim A\Lambda^{2}Z'.
\end{equation}
Because the energy of the domain wall configuration depends  linearly
on $Z'$, the direction of $Z'$ is not flat but stabilized
in the presence of  the domain wall configuration.
Here we neglect other effects , such as one-loop corrections to
the K\"ahler potential because
they will be very small in the $\lambda \rightarrow 0$ limit.
(See eq.(\ref{scalar}).)
\section{Cosmological implications of the domain wall}
\hspace*{\parindent}
In the last section we considered the domain wall configuration
for the vectorlike sector in the limit $\lambda \rightarrow 0$.
In this section, we will consider some specific examples and discuss
their astrophysical implications.
The first example is the dynamical supersymmetry breaking model
of ref.\cite{vector}.
This model contains two kinds of singlets which are assumed to
be stabilized by the effective K\"ahler
potential of the form:
\begin{equation}
  K=ZZ^{*}-\frac{\eta}{4\Lambda^{2}}\lambda^{4}(ZZ^{*})^{2}+...,
\end{equation}
where $\eta$ is a real constant and the order of the coupling constant
$\lambda$ is assumed to be O(1).
The effective scalar potential of the scalar $Z$ is given by
\begin{equation}
  \label{scalar}
  V_{eff}\sim \lambda^{2}\Lambda^{4}\left(1+\frac{\eta}{\Lambda^{2}}
    \lambda^{4}ZZ^{*}+...\right).
\end{equation}
The sign of $\eta$ is responsible for the vacuum expectation value of
$Z$.
When $\eta$ is positive, the above effective potential leads to
$<Z>=0$.
Otherwise, when $\eta$ is negative, $<Z>$ may be the order of $\Lambda$.
In ref.\cite{vector} it is assumed that one of the singlets
is stabilized at the origin but the other is stabilized at
$\Lambda \sim 10^{7}$Gev.
For this type of potential, domain walls can be produced during inflation,
reheating or the parametric resonance\cite{PR}.
It should be important to study the phenomenology of this domain
wall production\cite{new}.
The second example is the dynamical unification model of ref.\cite{dvali}.
The tree level superpotential is given by
\begin{equation}
  W_{tree}=gS trM_{ij} +\frac{g'}{2}STr \Sigma^{2}+\frac{h}{3}Tr \Sigma^{3}
\end{equation}
where $S$ is a singlet and $\Sigma$ is an adjoint of SU(5) GUT gauge
group.
Here $M_{ij}$ denotes
$M_{ij}\equiv Q_{i}\overline{Q_{j}}$
where $Q$ is not charged under SU(5) but in the fundamental representation
of the vectorlike gauge sector SU($N_{c}$).
The resulting dynamical superpotential which arises when we consider
the full quantum moduli space of the vectorlike sector is:
\begin{equation}
  W_{dyn}=A(detM_{ij}-B\overline{B}-\Lambda^{4})+
  S\left(g trM_{ij} +\frac{g'}{2}Tr \Sigma^{2}\right)+\frac{h}{3}
  Tr \Sigma^{3}.
\end{equation}
The minimum of this dynamical superpotential does not break
global supersymmetry.
SU(5) gauge symmetry is broken to SU(3)$\times$ SU(2)$\times$ U(1)
at the minimum:
\begin{eqnarray}
  <\Sigma>=\Lambda\sqrt{\frac{-2g}{15g'}}(2,2,2,-3,-3),
  &M_{i}^{j}=\delta^{j}_{i}\Lambda^{2},\nonumber\\
  B=\overline{B}=0,S=\Lambda h \sqrt{\frac{-2g}{15}}g'^{-3/2},
  & A=\Lambda^{-1}h
  \sqrt{\frac{2}{15}}\left(\frac{-g}{g'}\right)^{3/2}
\end{eqnarray}
where the scale of $\Lambda$ is fixed at $\sim 10^{15} GeV$.
A problem arises when we include supergravity effects.
Because the scale of gaugino condensation is about the
order of $gS\Lambda^{2}$, we should fine-tune the coupling
constant $h<10^{-9}$ so as not to break supersymmetry too much
by gaugino condensation.
With this fine-tuning, the vacuum expectation value of the singlet
$S$ is estimated to be $\sim 10^{6}$GeV and the height of the potential is
$V(0)^{1/4}\sim 10^{15}$GeV.
In this model, the domain wall production rate will be very small
because the potential barrier is very high.
\section{Conclusion and discussions}
\hspace*{\parindent}
In this paper we constructed the  domain wall solution for
ordinary $N_{c}=N_{f}$ supersymmetric QCD and extend it for
the vectorlike dynamical supersymmetry breaking model.
Astrophysical implications of the domain wall configuration
is also considered for two types of vectorlike sectors.
We think it is very important to study the dynamics of the moduli dependent
constants (gauge coupling\cite{matsuda} or mass parameter\cite{new})
in the early universe.
They may be displaced by fluctuations of any type and can trigger
rich phenomena.
\section{Acknowledgment}
It is a pleasure for me to express my gratitude to S.Iso for
enlightening discussions on physics of domain wall.

\end{document}